# Deep Convolutional Neural Networks Model-based Brain Tumor Detection in Brain MRI Images

Md. Abu Bakr Siddique
*Department of EEE*
*International University of Business Agriculture and Technology*
Dhaka, Bangladesh
absiddique@iubat.edu

Shadman Sakib
*Department of EECS*
*University of Hyogo*
Himeji, Japan
sakibshadman15@gmail.com
shadman.sakib@ieee.org

Mohammad Mahmudur Rahman Khan
*Department of ECE*
*Vanderbilt University*
Nashville, United States
mohammad.mahmudur.rahman.khan@vanderbilt.edu

Abyaz Kader Tanzeem
*Department of EEE*
*BRAC University*
Dhaka, Bangladesh
tanzeemabyaz@gmail.com

Madiha Chowdhury
*Department of URP*
*Bangladesh University of Engineering and Technology*
Dhaka, Bangladesh
madihac940@gmail.com

Nowrin Yasmin
*Department of CSE*
*Ahsanullah University of Science and Technology*
Dhaka, Bangladesh
nowrin_yasmin@outlook.com

*Abstract*—Diagnosing Brain Tumor with the aid of Magnetic Resonance Imaging (MRI) has gained enormous prominence over the years primarily in the field of medical science. Detection and/or partitioning of brain tumors solely with the aid of MR imaging is achieved at the cost of immense time and effort and demands a lot of expertise from engaged personnel. This substantiates the necessity of fabricating an autonomous model brain tumor diagnosis. Our work involves the implementation of a deep convolutional neural network (DCNN) for diagnosing brain tumor from MR images. The dataset, used in this paper, consists of 253 brain MR images where 155 images are reported to have tumors. Our model can single out the MR images with tumors with an overall accuracy of 96%. The model outperformed the existing conventional methods for the diagnosis of brain tumor in the test dataset (Precision = 0.93, Sensitivity = 1.00, and F1-score = 0.97). Moreover, the average precision-recall score of the proposed model is 0.93, Cohen's Kappa 0.91, and AUC 0.95. Therefore, the proposed model can be helpful for clinical experts to verify whether the patient has a brain tumor and, consequently, accelerate the treatment procedure.

## I. INTRODUCTION

A tumor results from an uncontrolled division of abnormal cells forming a mass that can hamper the normal functionality of the tissue or organ [1]. Tumors are distinguished with their origin as basis and cell types. The brain manifests early stages of tumor mostly in the cerebrum area, whereas secondary ones (metastatic) find its way to the brain from other parts of the body [2]. Tumors may be malignant (high-grade) which are cancerous or benign (low-grade). Compared to a benign brain tumor, a malignant brain tumor grows very rapidly and is more prone to invade adjacent tissues. Thus, a primary malignant brain tumor has a dreary prognosis and considerably reduces cognitive function and quality of life [3].

Cancers pertaining to the brain and various nervous system rank tenth in the leading causes of mortality and is considered as the 3$^{rd}$ most prevalent cancer among teenagers and adults [4]. The five-year survival rate for male and female patients with a cancerous brain is 34% and 36% respectively. Gliomas, the most prevalent cerebral tumor type in adults, account for almost 80% of malignant cases [2], [3]. Patients with low-grade gliomas (LGG) are found to have an overall ten-year survival rate of about 57% [5]. The causes of a brain tumor can be attributed to environmental aspects such as excessive usage of artificial chemicals or genetic factors. Treatment options include radiotherapy, chemotherapy, and surgical procedure.

The earlier a brain tumor is detected the higher the chances of survival and the wider the treatment options. Various methods may be employed to diagnose a brain tumor such as MRI scan, BIOPSY and SPECT (Single Photon Emission Computed Tomography) scan. Magnetic Resonance Imaging (MRI) is the most prevalent method due to its non-intrusive imaging modality that provides distinctive tissue contrast. It is also highly conformable in terms of normalization of tissue contrast providing minute details of interest. However, limitations are faced because of the uncertain, random, and irregular size, shape, and locality of brain tumors. Also, non-autonomous partitioning of the tumors comes at the cost of a lot of time and is a labor-intensive, cumbersome, and largely subjective task given the amount of data to be handled thus reducing accuracy.

The algorithms applied for segmentation of brain tumor can be categorized as traditional or non-autonomous techniques and techniques pertaining to deep-learning. The former includes regularized non-negative matrix factorization (NMF), Computer-aided Diagnosis (CAD) systems involving computation methods such as K-means clustering and Principal Component Analysis (PCA) and Support Vector Machines (SVM) [6], [7]. Referencing cases scanned by T1-weighted and T2-weighted sequences, Hsieh *et al.* [8] fuzzy clustering was coupled with region-growing which resulted in a 73% segmentation accuracy. Albarracín *et al.* [9] evaluated various clustering algorithms for brain tumor segmentation revealing that k-means, fuzzy k-means, and Gaussian mixture models were inferior to 77% accuracy achieved by the Gaussian hidden Markov random field algorithm. Soltaninejad *et al.* [10] graded brain tumors using SVM as a basis by utilizing aspects of 38 first-order or second-order statistical measurements. Recently, the deep learning (DL) based techniques developed are enticing to researchers as they are automated using a complex set of features directly from data [11].

In the study [12], tumor segmentation was performed using the LinkNet network. At first, all seven training datasets were segmented using s single LinkNet network.

Later, a method for CNN was developed for the autonomous extraction of most prevalent tumor types with no preprocessing requirements. The dice score obtained for a single network was 0.73 and multiple systems were 0.79. The study [13], also use multiple convolutional neural networks for brain tumor segmentation by inputting the 3D MRI image voxels into a two dimensional CNN model. The output is a brain tumor segmentation, not a classification. For grading brain tumors, Ahmmed *et al*. [14] used SVMs combined with Artificial Neural Networks (ANNs). ANNs reduced the error of the learning process resulting in an accuracy of 97.44%. However, the number of images used in this method for the training process is relatively less raising acceptability queries for datasets with a large number of images. Le *et al*. [15] performed studies on the ImageNet dataset. Using the concepts of CNNs, they came up with a high-performance object detector which achieved a 70% relatively improved accuracy compared to earlier researches on the ImageNet dataset.

Here, a proposition involving a deep convolutional neural network (DCNN) based on a pre-trained network is made. VGG-16 classification model pre-trained on the ImageNet dataset is utilized. The dataset employed is constituted of two classes of brain MRI images. The model yields better results than conventional methods. The overall article is laid out in the following manner: Section II elaborates details of the methodology propositioned. The experimental results, performance evaluation, and the result comparison with the conventional methods are outlined in Section III. Finally, section IV pertains to the conclusion.

## II. PROPOSED METHODOLOGY

The study undertakes the task of automatic detection of brain tumors in brain MRI images. The workflow of the proposed approach is illustrated in Figure 2. Our proposed approach makes use of DCNN architecture as the basis for brain tumor detection using brain MRI images. The proposed method consists of several steps. At first, the brain MRI image is taken as the input image. Next, data normalization is conducted where image thresholding and dilations have been applied dispense with noise. The assembled database of MRI images of the brain is processed and augmented. After that the images were resized for the model's input and a pre-trained CNN, VGG-16 is employed to classify the images into two classes of *YES* and *NO*. VGG-16 is a VGGNet version trained on the ImageNet platform and is one of the state-of-the-art networks used as classifiers for images.

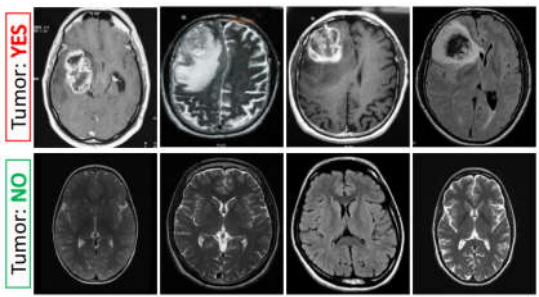

Fig. 1. Brain MRI images dataset sample

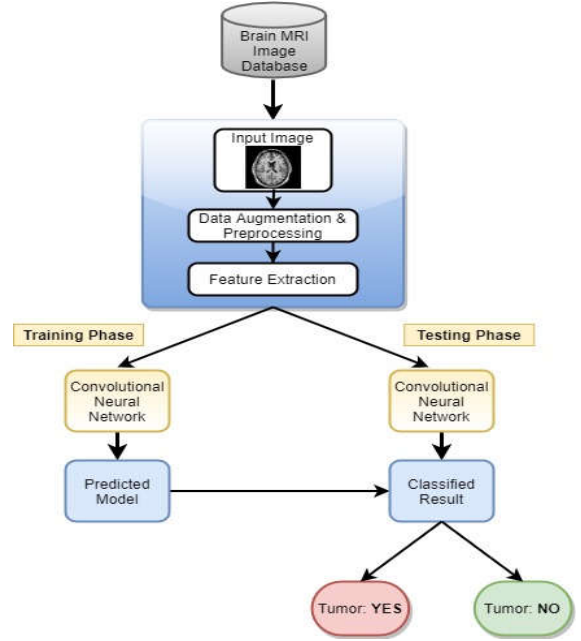

Fig. 2. The workflow of the proposed method

The database used in this study is composed of images of brain MRI scans. There are a total of 253 raw images of varying dimensions. The images are collected from Kaggle datasets of Brain MRI Images [16]. They are in JPG format. The dataset is labeled into two classes of YES and NO based on the presence of tumors. Overall, there are 155 images with brain tumors and the remaining 98 images are of normal brains. Figure 1 illustrates the sample images of the dataset.

### A. Preprocessing and Dataset Splitting

MR images could accumulate anomalies or, such as inhomogeneity distortions and heterogeneity of motion, due to motions produced by the subject while acquisition or MRI machine limitations. These artifacts trigger the induction of false intensity rates which lead to false positives in the image. Bias field distortion corrected by N4ITK system is used for working with these artifacts. Firstly, subsequent to inputting the MRI image, preprocessing was employed. As the intensity values in MR images are having different black corners, it is challenging for CNN to acclimate to the characteristics of the individual class label. Therefore, to enhance contrast intensity normalization is applied to narrow down intensity values to a stable range thus mean intensity value tends to zero and standard deviation inclines towards one. Initially, the images are thresholded with a threshold value of 45 to and a series of erosions and dilations are applied to remove any small regions of noise. Next, the images are normalized by capturing the largest contour of each image and cropping the images according to the top, bottom, left, and right extreme points of the contour. The overall normalization process is demonstrated in Figure 3. The output $i_0$ is given by,

$$i_0 = \frac{i - \mu_i}{\sigma_i} \qquad (1)$$

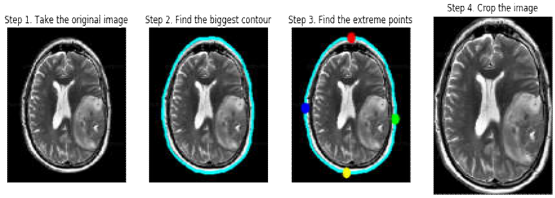

Fig. 3. Brain MRI images dataset sample

Here, $i$ indicates a normalized version of input MRI image, $\mu_i$ is the mean and $\sigma_i$ is the standard deviation.

The dimension pertaining to the input layer of the VGG-16 network is 224 × 224. Thus, to resize our dataset and make it fit for classification, data augmentation is employed. It is a process widespread throughout DL which helps in the generation of samples required. It also enhances the effectiveness of the network for a small database by optimizing it. Data augmentation plays the role of performance enhancement of the network by deliberately generating additional training data with the aid of original data. The images are augmented to create variation in them to overcome the size of the dataset being considerably small. Therefore, to improve the variance within our limited dataset, image augmentations are applied by grabbing Keras ImageDataGenerator while training. The entire augmentation step is as follows: the images are rotated randomly to 15° (clockwise). They are also shifted to 10% of the value of their height and width. The images are also randomly brightened or darkened to a range of 0 to 50%. Shear with an angle of 0.1 radians (counter-clockwise) is also applied to the images. Finally, the images are randomly flipped horizontally and vertically. The resized sample images for training the model is shown in Figure 4.

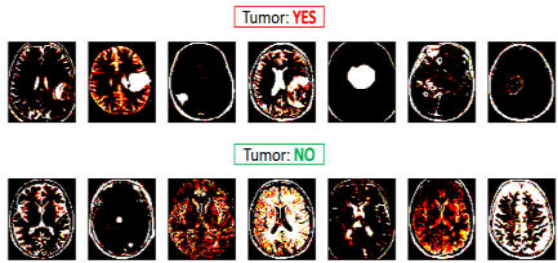

Fig. 4. Resized MRI images of the brain

The database is divided between training, testing, and validation sets with an 80:10:10 ratio. The 253 image dataset is split into training, testing, and validation sets of 207, 24, and 22 images respectively.

### B. Convolutional Neural Networks (CNN): VGG16

Convolutional Neural Networks (CNN) is a type of neural network that can extract important features from images, analyze those features, and classify them accurately. This allows CNN to be more suitable than any other conventional deep learning models in the field of image classification. There are many different architectures of CNN, but a general CNN architecture is constituted of 3 primary layers; the convolution layer which is followed by the pooling layer and the fully connected layer. The convolutional layers that work on a localized point, and not all places. This separates the distinctions from the original data, then converts that layer preceding it into the layer. After that, the pooling layer begins to learn from the preceding layer and moves towards reducing the difficulty of the operation. Finally, the fully connected layer executes the features acquired from all previous layers that provide the classified outputs required.

The VGG-16 is used as the extraction of features in our study. VGG-16 is a very renowned CNN based model and trained beforehand on extensive datasets like the ImageNet using at least one million images and is therefore very systematic and productive in the field of image classification. The architecture of VGG-16 shown in Figure 5, was first proposed by Simonyan and Zisserman [17] who were part of the Visual Geometry Group during the ImageNet Competition in 2014; their submissions secured them first and second places in the field of classification and localization respectively.

The VGG-16 neural network is trained with the aid of million labeled images which enables it to classify images into 1000 different classes. Its architecture consists of 41 layers in total, of which 16 have learnable parameters; 13 convolution layers coupled with 3 fully connected layers along with a rectified linear unit (ReLU) equipped with the hidden layers. The general model of VGG-16 has convolution layers that consist of 3 × 3 filters (small receptive field) with convolution stride and padding sizes of 1 pixel respectively. Subsequent to every convolution layer is a max-pooling layer.

VGG-16 accepts input images with size 224 × 224 which are relayed to the first two convolution layers. The first two convolution layers contain 64 feature kernel filters with size 3 × 3 which have convolution stride and padding sizes of 1 pixel respectively. The resulting feature map has a new dimension of 224 × 224 × 64 which is relayed to a max-pooling layer constituting of stride size of 2 and 2 × 2 kernels. The max-pooling layer is performed over a space of 2 × 2 kernels with a stride size of 2 pixels; this operation reduces the spatial dimension of the feature map from the previous layer by half i.e., 112 ×112 × 128. After going through the max-pooling layer, the output is relayed to the 3rd and 4th convolution layers respectively that contain 124 feature kernel filters with 3 × 3 dimensions. Following the 3rd and 4th convolution layers, the output is then propagated through another max-pooling layer which is performed over a space 2 × 2 kernels and a stride size of 2 pixels to produce a feature map of dimension 56 × 56 × 256. The feature map is propagated through the 5th, 6th, and 7th convolution layer that has 256 feature kernel filters with size 3 × 3. These layers are followed by another max-pooling which is performed over a space 2 × 2 kernels and a stride size of 2 pixels layer to produce a feature map of dimension 28 × 28 × 512. The two sets of convolution layers that come after, 8th till 13th have filters of size 3 × 3 and 512 feature maps. They are followed by max-pooling layers which have a stride size of 1 pixel and are performed over a space of 2 × 2 kernels. The final set of the network that follows consists of 3 Fully-Connected (FC) layers constituting of a filter of dimension 3 × 3 and has ReLU activated units of 4096, 4096, and 1000 respectively.

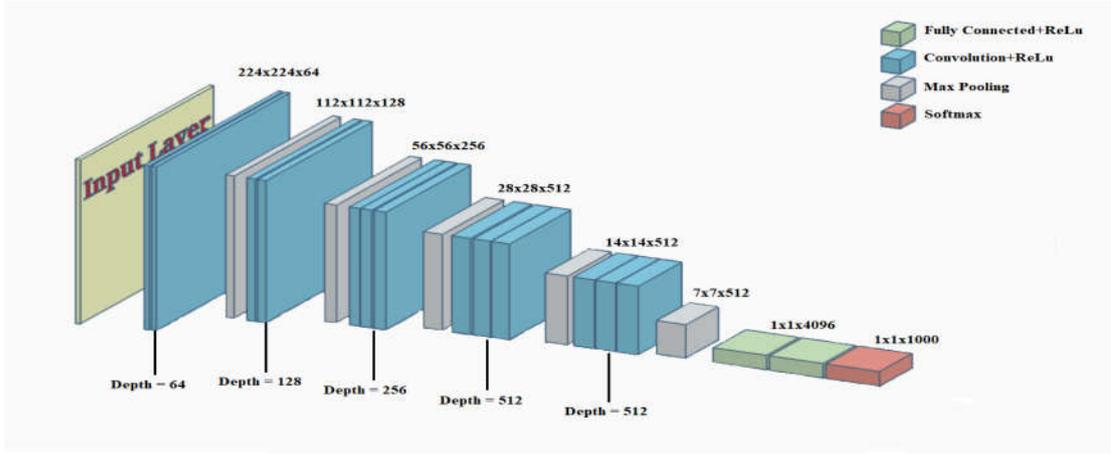

Fig. 5. General architecture of VGG16

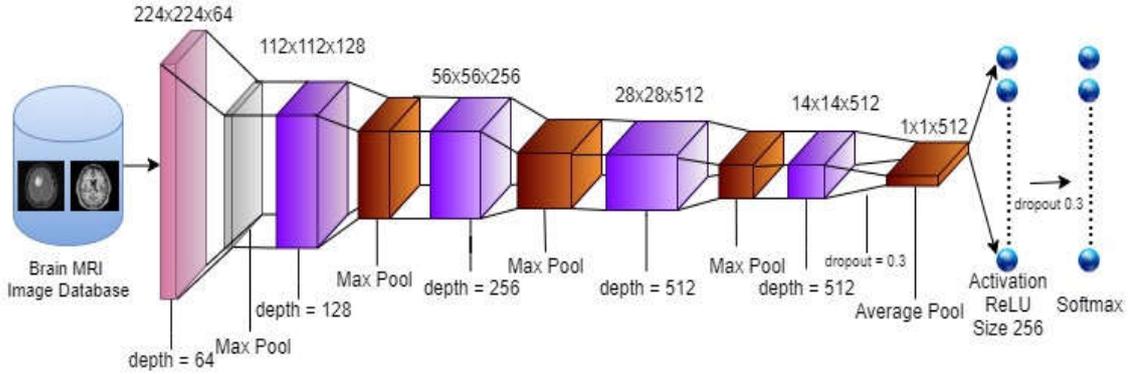

Fig. 6. The architecture of the proposed DCNN model

Mathematically, ReLU is expressed as,

$$f(x) = \max(0, x) \quad (2)$$

The Fully-Connected layers help in obtaining the feature vectors which are passed to the Softmax layer of 1000 units for classification. Equation 3 defines the Softmax activation function.

$$\sigma(z)_x = \frac{e^{z_x}}{\sum_i^l e^{z_i}} \quad (3)$$

### C. Proposed DCNN Model Architecture

In this paper, a pre-trained VGG-16 neural network model is employed for the classification that uses more than a million labeled images for its training obtained from large-scale datasets like the ImageNet. For our research, architecture has been modified. The proposed model as shown in Figure 6 is similar to the general VGG-16 model but the architecture has been modified to perform our task. In the proposed model, the final max-pooling layer from a general VGG-16 architecture is replaced by an average-pooling layer. The average-pooling layer, also known as Global Average Pooling Layer (GAP), performs spatial pooling of the feature map to reduce overfitting by limiting the overall amount of parameters in the model. However, the reduction in the spatial dimensions by average-pooling is more significant than that of max-pooling.

This replacement of the max-pool layer by a GAP layer proved to be effective in producing highly improved results. Moreover, the set of convolution layers in the center just before the fully connected layers have been frozen so that the weights of the layers cannot be modified while running the backpropagation algorithm. However, the GAP layer and the fully connected layers have been allowed for modification. Freezing the initial layers enhances the training time of the neural network by a great deal. In addition, to minimize overfitting and hence generalization error, two dropout values of 0.3 have been incorporated into between the GAP layer and the fully connected layers. Finally, fully connected layers that have 256 channels each and a Softmax layer that has 2 channels were used.

### III. RESULT ANALYSIS AND DISCUSSION

In our implemented fine-tuned DCNN, VGG-16, 80% of the total data was reserved as training data, 10% of it for validation, and the final 10% for testing. The neural network is trained by employing Adam optimizer as the optimization algorithm using a learning rate of 0.0001 for 80 epochs, the batch volume of 16, and categorical cross-entropy as the loss function. A computer system have been used with the Linux 4.0.0 LTS platform (Ubuntu 16.04.6), Geforce RTX 2080Ti GPU during our experimentation. The DCNN model was validated on intel Xeon-2620, Core i5-2.4GHz CPU, 16 GB RAM, utilizing python in the Keras module.

## A. Models Performance Evaluation

This segment covers the performance evaluation of employed and trained DCNN model. The aforementioned model proposed in this paper was implemented on the Brain MRI dataset which consists of 253 brain MR images where 155 images show indications of tumors. As the dataset is distorted, precision, sensitivity or recall, F1-score, average precision-recall score, and Cohen's Kappa measurement to clarify the supremacy of the proposed approach is interpreted, instead of merely focusing on the classification accuracy as a model performance evaluating metric. If a double class classification issue is considered, the performance of a classifier model can be laid out as a confusion matrix shown in Table I.

TABLE I. CONFUSION MATRIX [18]

|  | Expected Positive | Expected Negative |
|---|---|---|
| Actual Positive | TP | FN |
| Actual Negative | FP | TN |

The parameters TP symbolize the true positive for the number of positive brain tumor images properly classified, TN depicts the number of images that were classified correctly. Conversely, FP denotes false-positive images which is the number of positive images incorrectly classified as negative, and FN denotes the false-negative as represents the number of negative images misclassified as positive.

With reference to the confusion matrix provided in Table I, the overall accuracy can be determined with the help of Equation 4.

$$Accuracy(\%) = (TP + TN) / (TP + FP + TN + FN) \quad (4)$$

However, as mentioned earlier, the precision, sensitivity, or recall and the F1-score values for analyzing the DCNN model's performance is also evaluated. The equations of these metrics are provided in Equation 5, 6, 7, and 12, respectively. The precision value represents the classifier model's exactness. On the other hand, the recall value represents the model's completeness.

$$Precision(PPV) = TP / (TP + FP) \quad (5)$$

$$Sensitivity = TP / (TP + FN) \quad (6)$$

$$F1-score = \frac{2 \times (Precision \times Recall)}{Precision + Recall} \quad (7)$$

$$p_0 = \frac{TP + TN}{TP + FN + FP + TN} \quad (8)$$

$$p_{yes} = \frac{TP + TN}{TP + FN + FP + TN} \cdot \frac{TP + FP}{TP + FN + FP + TN} \quad (9)$$

$$p_{no} = \frac{FP + TN}{TP + FN + FP + TN} \cdot \frac{FN + TN}{TP + FN + FP + TN} \quad (10)$$

$$p_e = p_{yes} + p_{no} \quad (11)$$

$$Cohen's\ Kappa = \frac{p_0 - p_e}{1 - p_e} \quad (12)$$

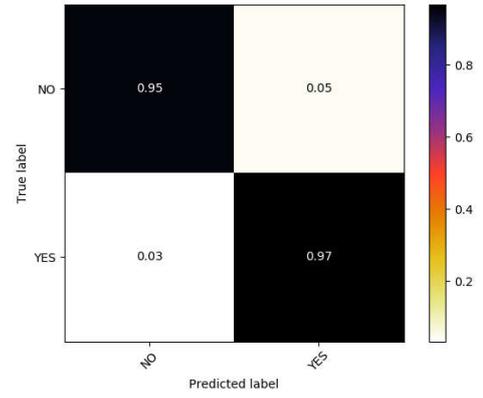

Fig. 7. Normalized confusion matrix on test data

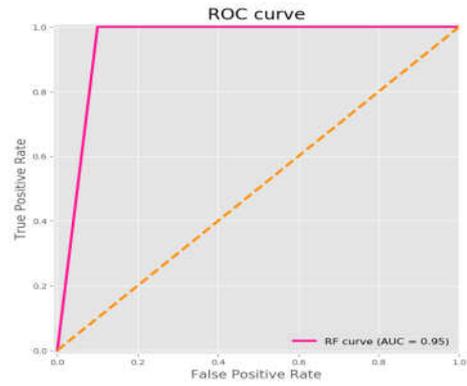

Fig. 8. ROC curve of the model propositioned

The corresponding values of the performance evaluating metrics are provided in Table II. The proposed DCNN approach works well to identify the brain tumor in MR images. Apart from demonstrating higher classification accuracy (96%), the model showed a high F1-score (0.97) and higher precision-recall values (0.93) which are the indications of the effectiveness of the model.

TABLE II. PERFORMANCE OF THE PROPOSED DCNN MODEL

| Performance evaluating metrics | Performance score |
|---|---|
| Precision | 0.93 |
| Sensitivity or Recall | 1.00 |
| F1-score | 0.97 |
| Average precision-recall score | 0.93 |
| Cohen's Kappa | 0.91 |
| AUC | 0.95 |
| Accuracy | 0.96 |

The ROC curve and the confusion matrix are provided in Figure 7 and Figure 8, respectively. The ROC curve suggests that the model has a higher ability to distinguish between the MR images with tumors and the images without a tumor with a high AUC score (95%). As seen from the slope of the confusion matrix that the *YES* class achieved the highest classification accuracy of 97% in detecting the brain tumor. So, the proposed DCNN model in this paper can be a suitable method for faster and effective brain tumor detection.

## B. Performance comparison with the existing methods

To assess the performance of the model, our proposed DCNN classifier was compared with conventional approaches. Table III displays the performance comparison of the model, put forward with other developed algorithms. Our proposed model outperforms the previously developed approaches demonstrating accuracy of 96%.

TABLE III. PERFORMANCE OF THE PROPOSED DCNN MODEL

| Method | Algorithms | Classification Accuracy |
|---|---|---|
| Amin et al. [19] | 7 layered 2D CNN | 95.1% |
| Reza et al. [20] | MFDFA + random Forest | 86.7% |
| Hemanth et al. [21] | LinkNet | 91.0% |
| Mohsen et al. [22] | SMO + SVM | 93.9% |
| **Proposed** | **DCNN (VGG16)** | **96.0%** |

## IV. CONCLUSION

Detection of brain tumors has a very significant role to play in medical interventions. It plays an integral role in the processing of medical images, as medical images have different variations. With the automatic brain tumor detection method, diagnosis is not only intuitive but also strengthens enormously the possibility of the patient's survival. The use of CNN models for the diagnosis of brain tumors has led the way to improve the accuracy of tumor diagnosis and classification. MRI is used most widely for the detection and classification of brain tumors. In this research, a deep neural convolution network (DCNN) is developed to automatically detect the brain tumor from obtained MRI images of the brain. The model was trained for faster and effective training using a pre-trained VGG-16 network. Results showed that the proposed architecture of the network is tempting and operates particularly well in detecting tumors in contrast to conventional methods. Several preprocessing procedures were conducted to improve the model's efficiency. Results validate that the model achieved an overall classification accuracy of 96%, Precision of 0.93, Sensitivity 1.00, and F1-score of 0.97. Moreover, the model scored AUC (95%), average precision-recall score (0.93), and Cohen's Kappa 0.91. Thus, the proposed framework could be executed as a handy system for doctors to provide desirable medical treatments for early detection of brain tumors. To dictate the precise location of the tumor utilizing CNN, our work is expected to expand with 3D brain scans in the future.